\documentclass[useAMS,usenatbib]{mn2e}
\usepackage{amsmath}
\usepackage{epsfig}
\usepackage{txfonts}

\usepackage{natbib}
\usepackage[pdftex,pdfpagemode={UseOutlines},bookmarks,bookmarksopen,colorlinks,linkcolor={blue},citecolor={blue},urlcolor={red}]{hyperref}
\usepackage{float}
\usepackage{graphicx}
\usepackage{caption}
\usepackage{subcaption}
\usepackage{natbib}
\usepackage{rotating,longtable,lscape}
\usepackage{multirow}
\usepackage{pifont}
\usepackage{etoolbox}
\makeatletter
\patchcmd\@combinedblfloats{\box\@outputbox}{\unvbox\@outputbox}{}{%
  \errmessage{\noexpand\@combinedblfloats could not be patched}%
}%
\makeatother
\newcommand{\cmark}{\ding{51}}%
\title[CBP host mass ratios]{The binary mass ratios of circumbinary planet hosts}

\author[Martin]
{\parbox{\textwidth}{David. V. Martin}
\vspace{0.4cm}\\
\parbox{\textwidth}{Department of Astronomy and Astrophysics, University of Chicago, 5640 South Ellis Avenue, Chicago, IL 60637, USA \\
Fellow the Swiss National Science Foundation
\\ davidmartin@uchicago.edu}}

\begin{document}

\date{Submitted 28-05-2018, Revised 28-02-2019, Accepted 25-03-2019}

\pagerange{\pageref{firstpage}--\pageref{lastpage}} \pubyear{2018}

\maketitle

\label{firstpage}

\begin{abstract}

Almost a dozen circumbinary planets have been found transiting eclipsing binaries. For the first time the observational bias of this sample is calculated with respect to the mass ratio of the host binaries. It is shown that the mass ratio affects transit detection in multiple, sometimes subtle ways, through stability and dynamics of orbits, dilution of transit depths and the geometric transit and eclipse probabilities. Surprisingly though, it is found that these effects largely cancel out. Consequently, the transit detections in the {\it Kepler} mission are essentially unbiased with respect to mass ratio, and hence likely representative of the true underlying population. It is shown the mass ratio distribution of circumbinary hosts may be the same as field binaries, and hence roughly uniform, but more observations are needed to deduce any subtle differences. These results are discussed in the context of close binary formation and evolution, of which the mass ratio is believed to be a marker, and other surveys for circumbinary planets including {\it TESS} and BEBOP.



\end{abstract}


\section{Introduction}
\label{sec:intro}


The {\it Kepler} space telescope precipitated the discovery of 11 transiting circumbinary planets. Owing to a three-body geometry, a dynamically-varying planet orbit and constraints of orbital stability, the detection biases are more complicated than for planets transiting single stars. Understanding these biases though is essential for uncovering the underlying population. The current sample, whilst small in size, has yielded preliminary insights on the occurrence rate of circumbinary planets \citep{Armstrong:2014yq}, their orbital architectures \citep{Martin:2014lr,Li:2016ng} and the orbital periods of their host binaries \citep{Munoz:2015uq,Martin:2015iu,Hamers:2016er,Xu:2016qw,Fleming:2018mf}.

A property which is yet to be considered is the binary mass ratio: $q=M_{\rm B}/M_{\rm A}$, where A and B refer to the primary and secondary star, respectively. Whilst these mass ratios were published in the discovery papers, they have never been analysed as an ensemble. Planets have been found transiting binaries spanning almost all possible binary mass ratios, including a slight over abundance at small $q\in[0.2,0.35]$. However, without knowing how $q$ biases the transit detection, we cannot deduce any connections between the mass ratio and planet occurrence. Such a question is interesting in the context of tight binary formation. The favoured theory is an inwards migration from a primordially wide orbit \citep{Bonnell:1994ht,Bodenheimer:1995aq,Simon:1995mq,Larson:2002bg}. This process is already thought to affect the binary mass ratio, but its effect on circumbinary planets is yet to be determined.

In this paper we calculate the detectability of circumbinary planets as a function of $q$, including the transit geometry, stability limit, evolution of the planetary orbit and dilution of transit depths. We de-bias the observed {\it Kepler} distribution of circumbinary planet hosts. and compare it to field binaries of different periods. This comparison has implications for how the binary evolution process may affect, or possibly be affected by the presence of a circumbinary planet.

This work is applicable to any long-base line transit survey, so we naturally apply it to the original {\it Kepler} mission. However, it is also relevant for the year-long {\it TESS} observations of the ecliptic poles, and the future {\it PLATO} mission.

The organisation of this paper is as follows. First, in Sect.~\ref{sec:kepler_CBPs} we present the circumbinary discoveries to date. Then, in Sect.~\ref{sec:effects_and_biases} we outline all of the different selection effects that bias transit observations. In Sect.~\ref{sec:de-biasing} we take these selection effects and deduce the overall observational bias as a function of $q$, using simulations of various circumbinary distributions. Using this, in Sect.~\ref{sec:observations} we de-bias the observed distribution of $q$ and compare it to mass ratios of binaries discovered in various other observational surveys. Using these results, in Sect.~\ref{sec:discussion} we discuss implications for both planet and binary formation and motivate new research, both theoretical and observational. We also consider some limitations of our work. Following this is a brief conclusion in Sect.~\ref{sec:conclusion}.

\section{The {\it Kepler} circumbinary planets}\label{sec:kepler_CBPs}


So far 11 planets have been discovered orbiting 9 eclipsing binaries. Some basic parameters of these systems are listed in Table~\ref{tab:KeplerDiscoveries}. There is one multi-planet system - Kepler-47 - which contains three. In all of the discoveries the planet transits the primary star. However, in only four of the nine binaries are secondary transits detected. All cases without secondary transits have a small mass ratio: $q<0.5$. It is expected that secondary transits did geometrically occur for some of these systems \citep{Martin:2017qf}, but the relative faintness of the secondary star made them elude detection. The planets are all larger than $\sim 3R_{\oplus}$, although there is a detection difficulty which biases us against smaller planets \citep{Armstrong:2014yq,Martin:2018gr}. The planets also typically orbit with periods $\sim 5-6$ times that of the binary. This places them close to the dynamical stability limit (\citealt{Martin:2014lr,Li:2016ng,Quarles:2018ub} and see Sect.~\ref{subsec:stability}). Finally, all of the orbits are within $\sim 4^{\circ}$ of coplanarity \citep{Martin:2014lr,Li:2016ng}.




\begin{table*}
\caption{Orbital parameters of the transiting systems discovered so far by {\it Kepler}.} 
\centering 
\begin{tabular}{ cccccccccc } 
\hline\hline 
{\it Kepler} & $M_{\rm A}$ & $M_{\rm B}$ & $q$ & $R_{\rm A}$ & $R_{\rm B}$ & $T_{\rm bin}$ & $T_{\rm p}$ & primary & secondary \\ 
[0.5ex] 
number& ($M_{\odot}$) & ($M_{\odot}$) & & ($R_{\odot}$) & ($R_{\odot}$) & (day) & (day) & transits &transits \\
\hline
\hline
16 & 0.690 & 0.203 & 0.290 & 0.649 & 0.226   & 41.079 & 228.776 & \cmark & \cmark \\
\hline
34 & 1.048 & 1.021 & 0.971 & 1.162 & 1.093  & 27.796 & 288.822 & \cmark & \cmark \\
\hline
35 & 0.888 & 0.809 & 0.910 & 1.028 & 0.786  & 20.734 & 131.458 & \cmark & \cmark \\
\hline
38 & 0.949 & 0.249 & 0.263 & 1.757 & 0.272  & 18.795 & 105.595 & \cmark & \\
\hline
47 & 1.043 & 0.362 & 0.346 &  0.964 & 0.351   & 7.448  & 49.514, 187.3, 303.158 & \cmark & \\
\hline
64 & 1.384 & 0.336 & 0.268 & 1.734 & 0.378   & 20.000 & 138.506 & \cmark &  \\
\hline
413 & 0.820 & 0.542 & 0.659 &  0.776 & 0.484   & 10.116 & 66.262 & \cmark & \\
\hline
453 & 0.934 & 0.194 & 0.204 & 0.833 & 0.214   & 27.322 & 240.503 & \cmark &\\
\hline
1647 & 1.221 & 0.968 & 0.795 & 1.790 & 0.966   & 11.259 & 1107.592 & \cmark & \cmark \\
\hline 

\end{tabular}
\label{tab:KeplerDiscoveries}
\end{table*}


%

\section{Observational biases and effects}\label{sec:effects_and_biases}

\subsection{Binary eclipse probability}\label{subsec:EB_catalog}

The transiting circumbinary planets have all been found around eclipsing binaries. One must therefore understand the geometric biases in the eclipsing binary distribution. The criterion for a circular binary to eclipse is 

\begin{equation}
\label{eq:EclipseCriterion}
\sin\left|\frac{\pi}{2}-I_{\rm bin}\right| \leq \frac{R_{\rm A}+R_{\rm B}}{a_{\rm bin}},
\end{equation}
 where we allow for grazing eclipses, which are typically detectable with an instrument as precise as {\it Kepler}. To re-write Eq.~\ref{eq:EclipseCriterion} as a function of $q$ we use the quadratic mass-radius relation of \citet{Eker:2018nw},

\begin{equation}
\label{eq:mass_radius_relation}
\frac{R}{R_{\odot}} = 0.438\left(\frac{M}{M_{\odot}}\right)^2 + 0.479\frac{M}{M_{\odot}} + 0.075,
\end{equation}
which has been calibrated over a mass range of $[0.179,1.5]M_{\odot}$. This corresponds to the known circumbinary planet hosts, with the slight exception of Kepler-64 where $M_{\rm A}=1.53M_{\odot}$. Substituting Eq.~\ref{eq:mass_radius_relation} into Eq.~\ref{eq:EclipseCriterion} yields

\begin{equation}
\label{eq:EclipseCriterion_MassRatio}
\sin\left|\frac{\pi}{2}-I_{\rm bin}\right| \leq \frac{0.438M_{\rm A}^2\left(1+q^2\right) + 0.479M_{\rm A}\left(1+q\right) + 0.15}{a_{\rm bin}}.
\end{equation}
This criterion is expectedly easier to fulfill for higher values of $q$.

Additional complications related  relating to the construction of the eclipsing binary catalog, such  evolved stars and the Malmquist bias, are discussed in Sect.~\ref{subsec:limitations}.

\subsection{Planet transit probability}\label{subsec:transit_probability}

\begin{figure}  
\begin{center}  
\includegraphics[width=0.5\textwidth]{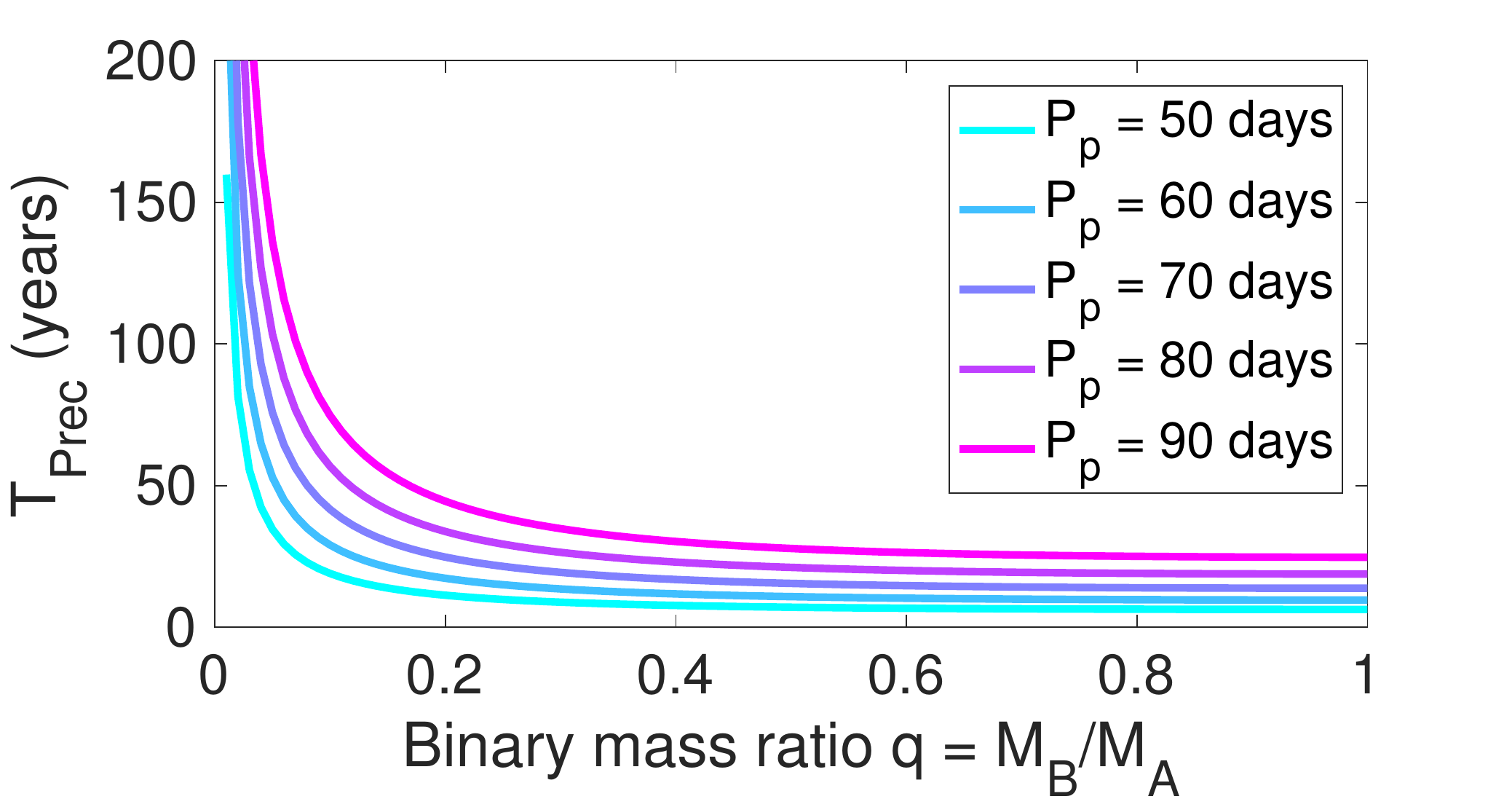}  
\caption{Precession period of a circumbinary planet of different periods around a binary with a 10 day period, a $1M_{\odot}$ primary star and binary mass ratio varying between 0 and 1.}
\label{fig:T_prec}
\end{center}  
\end{figure}

A key characteristic of circumbinary planets is that their orbit is not static; the substantial tidal potential of the binary induces a nodal precession. For circular orbits, with respect to the binary the planetary orbit circulates at a constant rate with a precession period

\begin{equation}
\label{eq:Tprec}
T_{\rm prec} = \frac{4}{3}\left(\frac{T_{\rm p}^7}{T_{\rm bin}^4} \right)^{1/3}\frac{\left(M_{\rm A} + M_{\rm B}\right)^2}{M_{\rm A}M_{\rm B}}\frac{1}{\cos \Delta I},
\end{equation}
\citep{Farago:2010fj}, whilst maintaining a constant mutual inclination $\Delta I$. Owing to the relatively small mass of the planet, the binary orbit can be considered static. Eq.~\ref{eq:Tprec} can be re-written as a function of the binary mass ratio $q$:
\begin{equation}
\label{eq:Tprec2}
T_{\rm prec} = \frac{4}{3}\left(\frac{T_{\rm p}^7}{T_{\rm bin}^4} \right)^{1/3}\frac{(1+q)^2}{q}\frac{1}{\cos \Delta I}.
\end{equation}
In Fig.~\ref{fig:T_prec} we plot $T_{\rm prec}$ as a function of the binary mass ratio. It is seen to be a reasonably flat function for a large range of mass ratios, with only a sharp increase as $q$ decreases below 0.2, as we move from the binary star domain to the star-planet domain.

The observational consequence of this precession is that the inclination of the planet on the plane of the sky follows a sinusoidal function charactised by:

\begin{equation}
\label{eq:Ip_over_time}
I_{\rm p}(t) = \Delta I \cos \left(\frac{2\pi}{T_{\rm prec}}\left(t-t_0 \right) \right) + I_{\rm bin}.
\end{equation}

A necessary but not sufficient condition for a planet to transit is that its orbit overlaps with that of the binary. \citet{Martin:2014lr,Martin:2015rf} define this as ``transitability''. For some configurations transits are not guaranteed on every passing of the binary, however those studies show that missed transits are only frequent for misalignments above $\gtrsim 5^{\circ}$, which is greater than all of the known circumbinary planets. Since we are only considering the known, near-coplanar systems, we will consider transitability windows to be equivalent to transit windows.

\citet{Martin:2017qf} calculates that the transit window corresponds to $I_{\rm P}$ within these limits around $90^{\circ}$:

\begin{multline}
\label{eq:Ip_transit_limits}
\left. I_{\rm p}\right|_{\rm transit} = \frac{\pi}{2} \pm \frac{1}{a_{\rm p}}\Bigg[R_{\rm A} + a_{\rm bin}\frac{q}{1+q}\frac{\cos \Delta I}{\sin I_{\rm bin}} \Bigg. \\ 
\Bigg. \times \sqrt{\tan^2\left(\cos^{-1}\left[\frac{\cos \Delta I}{\sin I_{\rm bin}} \right] \right) + \cos^2 I_{\rm bin}}  \quad \Bigg].
\end{multline}
where the equations have been re-written to be a function of $q$. The time of overlap is is calculated from solving Eq.~\ref{eq:Ip_over_time} for $t$ using Eqs.~\ref{eq:Ip_transit_limits}. Depending on the parameters, there may be zero, one or two regions of transitability within a precession period, and hence zero, two or four times $t$ to solve for. \citet{Martin:2017qf} also derive similar equation for the secondary star, but throughout this paper we consider transits of the primary star to be the criterion for detectability, since all 9 circumbinary systems have primary transits but only 4 have secondary transits. 

\subsection{Dynamical stability limit}\label{subsec:stability}

Circumbinary orbits are only stable if the planet remains sufficiently far from the binary. This field of three-body stability has been studied by many authors over the years \citep{Dvorak:1986fk,Holman:1999lr,Mardling:2001hj,Pilat-Lohinger:2003zx,Mudryk:2006po,Quarles:2018ub}. A rule of thumb is that $a_{\rm p}\gtrsim 3 a_{\rm bin}$ for stability \citep{Schneider:1994lr}. However, a more detailed investigation uncovers dependencies on both the binary and planet eccentricities and, relevant for this study, the binary mass ratio. 

The often-quoted study of \citet{Holman:1999lr} numerically derived a stability limit of

\begin{equation}
\begin{split}
\label{eq:stability_limit}
\frac{a_{\rm crit}}{a_{\rm bin}} &= 1.60 = 5.10 e_{\rm bin} - 2.22 e_{\rm bin}^2 + 4.12\frac{q}{1+q} - 4.27 e_{\rm bin}\frac{q}{1+q} \\ 
&- 5.09 \left(\frac{q}{1+q}\right)^2+ 4.61 e_{\rm bin}^2\left( \frac{q}{1+q} \right)^2.
\end{split}
\end{equation}
This equation has been re-written from \citet{Holman:1999lr} to use $q$, whereas their paper uses $\mu = M_{\rm B}/(M_{\rm A}+M_{\rm B})$, which is related to $q$ by $\mu = q/(1+q)$. Note that \citet{Holman:1999lr} assumed circular planetary orbits. Effects of the outer (planetary in our case) eccentricity were analysed for example by \citet{Mardling:2001hj}.

\begin{figure}  
\begin{center}  
\includegraphics[width=0.5\textwidth]{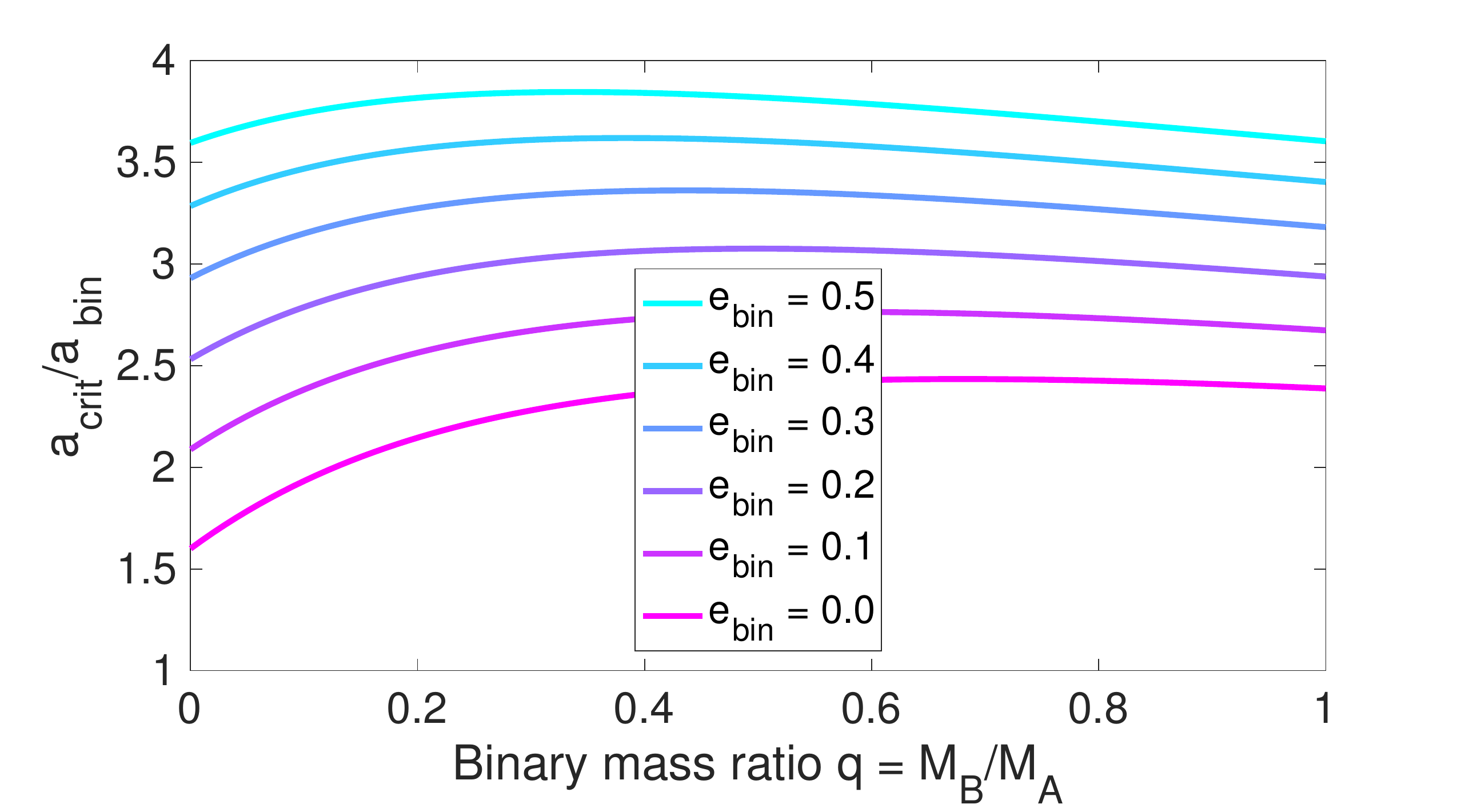}  
\caption{Stability limit from \citet{Holman:1999lr} calculated using Eq.~\ref{eq:stability_limit}. The range of binary eccentricities shown here corresponds to the range of applicability of their work.}
\label{fig:stability_limit}
\end{center}  
\end{figure} 

In Fig.~\ref{fig:stability_limit} we plot the stability limit as a function of $q$ for five different binary eccentricities, spanning the range of validity on which \citet{Holman:1999lr} derived their formula. The dependence on $q$ is shown to be weak. For $e_{\rm bin}=0.5$ the limit varies by as little as $7\%$ for all values of $q$, with it curiously having the same value for $q=0$ and $q=1$. For a circular binary there is a maximal variation of 34\%, with small mass ratios being more stable.

\subsection{Dilution of transit signals}\label{subsec:dilution}

%
%

If the light of a second star is mixed with the light of the star being transited, then the effect is to dilute the transit depth of the planet. This occurs regardless of whether the second star is bound or not, but merely requires it to be unresolved by the detector. If a circumbinary planet transits across the primary star A, then the observed transit depth will be:

\begin{equation}
\label{eq:transit_depth_binary}
\delta = \left(\frac{F_{\rm A}}{F_{\rm A} + F_{\rm B}}\right)\left(\frac{R_{\rm p}}{R_{\rm A}} \right)^2 = \left(\frac{M_{\rm A}^{3.5}}{M_{\rm A}^{3.5} + M_{\rm B}^{3.5}}\right)\left(\frac{R_{\rm p}}{R_{\rm A}} \right)^2,
\end{equation}
where $F_{\rm A}$ and $F_{\rm B}$ are the fluxes of the primary and secondary stars, respectively,  which are converted to mass according to the simple mass-luminosity of $L/L_{\odot} (M/M_{\odot})^{3.5}$ \citep{Kippenhahn:1990hr}. Re-arranging for $R_{\rm p}$ yields

\begin{equation}
R_{\rm p} = R_{\rm A}\sqrt{\delta \frac{M_{\rm A}^{3.5} + M_{\rm B}^{3.5}}{M_{\rm A}^{3.5}}}.
\end{equation}
Now substitute in the mass ratio $q=M_{\rm B}/M_{\rm A}$:
\begin{equation}
R_{\rm p} = R_{\rm A}\sqrt{\delta (1+q^{3.5})}.
\end{equation}

\begin{figure}  
\begin{center}  
\includegraphics[width=0.5\textwidth]{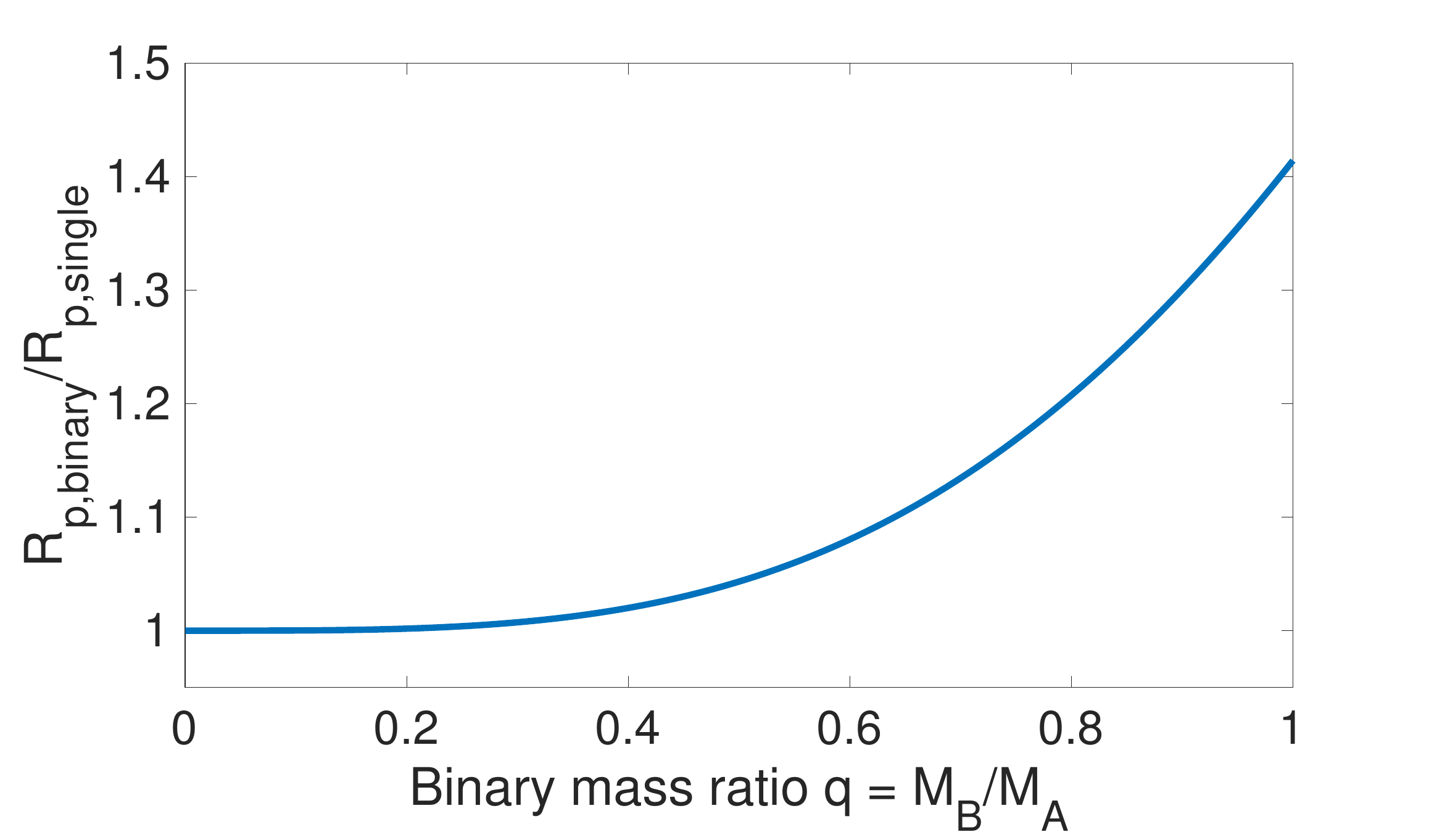}  
\caption{Ratio of a detectable planet radius transiting the primary star in a binary system of different mass ratios compared with a single star. Dilution from the secondary star reduces the transit depth, making planet detection more difficult. The governing Eq.~\ref{eq:R_detectable_ratio} is derived assuming a mass-luminosity relation of $L\propto M^{3.5}$.}
\label{fig:R_detectable_ratio}
\end{center}  
\end{figure}

From this equation we see the effect on the detectable planet radius as a function of the mass ratio $q$. Consider a given threshold of a detectable transit depth, which would be calculated primarily as a function of the magnitude of the star. The ratio of the smallest detectable planet around a single star verses one in a binary star system is:

\begin{equation}
\label{eq:R_detectable_ratio}
\frac{R_{\rm p, binary}}{R_{\rm p, single}} = \sqrt{1+q^{3.5}}.
\end{equation}
We plot Eq.~\ref{eq:R_detectable_ratio} in Fig.~\ref{fig:R_detectable_ratio}. For $q\lesssim0.4$ there is very little effect of dilution. The effects then become more pronounced at higher mass ratios, reaching an expected worst case dilution factor of $\sqrt{2}$. 

\section{Simulating the detection bias as a function of mass ratio}\label{sec:de-biasing}


We create a population of circumbinary systems (Sect.~\ref{subsec:de-biasing_initial}), simulate which planets will transit as a function of the various selection biases (Sect.~\ref{subsec:de-biasing_observed}) and then analyse the results (Sect.~\ref{subsec:de-biasing_results}).


\subsection{Initial population}\label{subsec:de-biasing_initial}

\begin{figure*}  
\begin{center}  
	\begin{subfigure}[b]{0.33\textwidth}
		\caption{Initial mass function}
		\vspace{-2ex}
		\includegraphics[width=\textwidth]{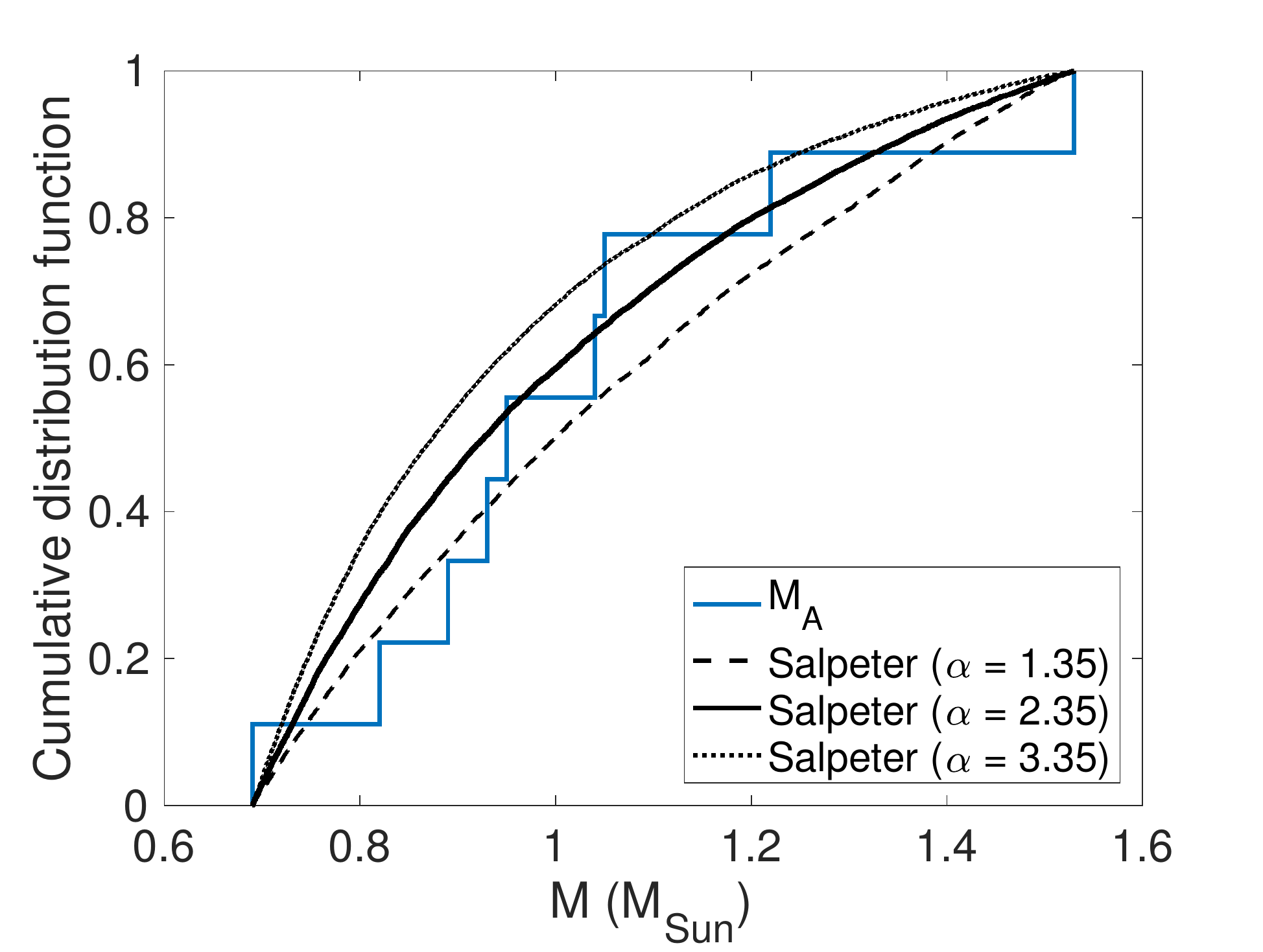}  
		\label{fig:imf}
		\vspace{3ex}
	\end{subfigure}
	\begin{subfigure}[b]{0.33\textwidth}
		\caption{Mass-radius relation}
		\vspace{-2ex}
		\includegraphics[width=\textwidth]{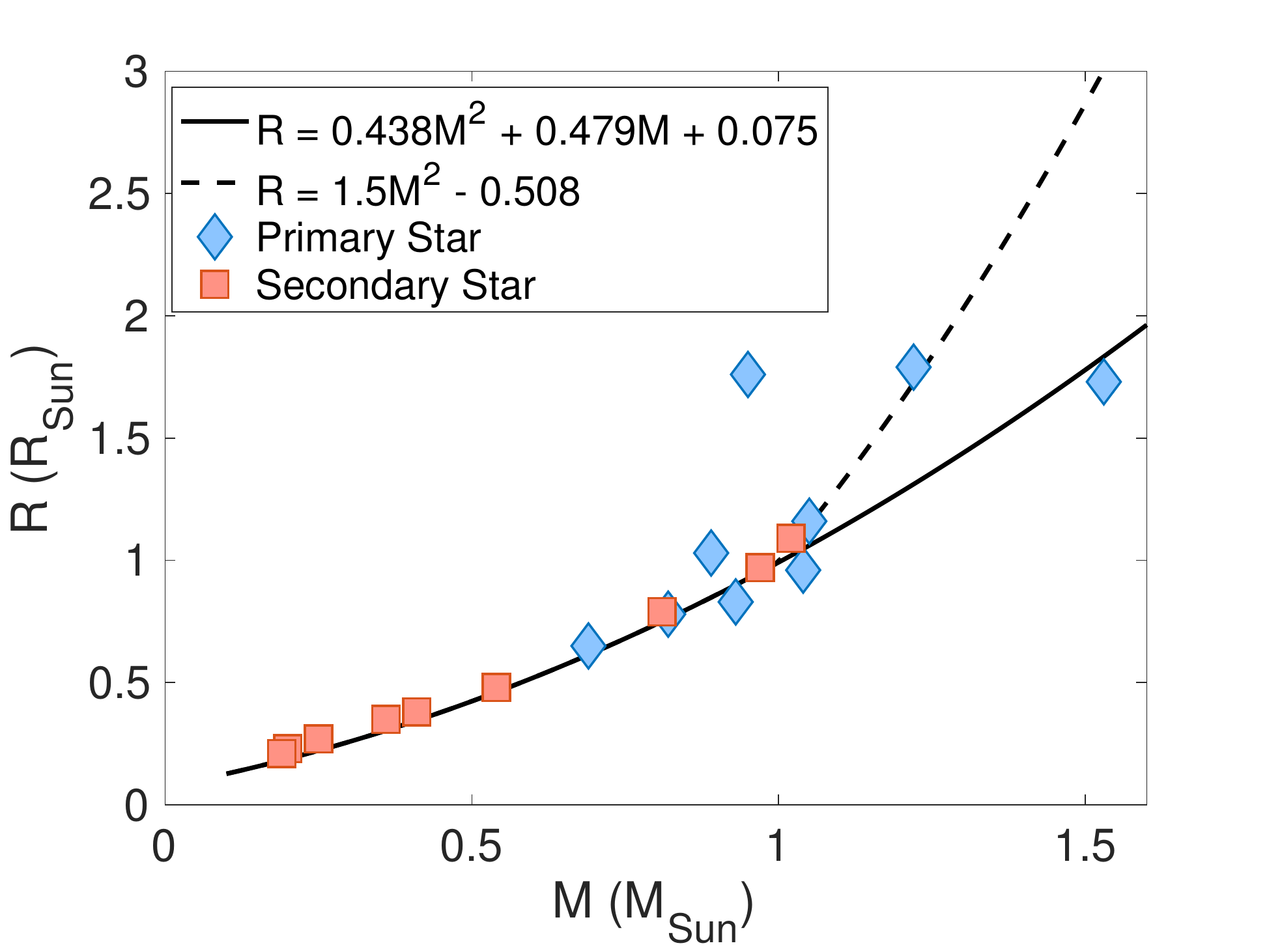}  
		\label{fig:mass_radius_relation}
		\vspace{3ex}
	\end{subfigure}
	\begin{subfigure}[b]{0.33\textwidth}
		\caption{Mutual inclination distribution}
		\vspace{-2ex}
		\includegraphics[width=\textwidth]{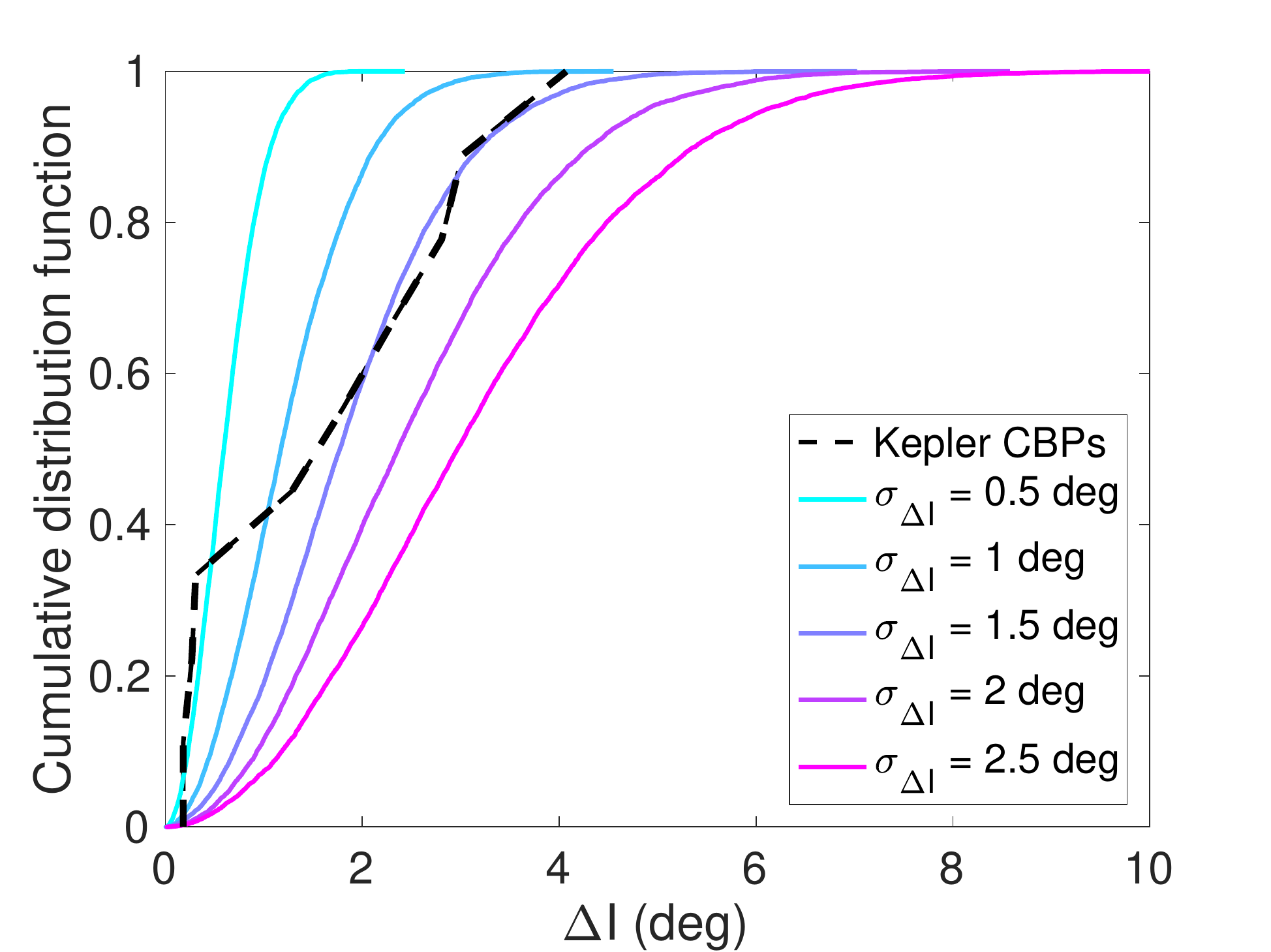}  
		\label{fig:DeltaI}
		\vspace{3ex}
	\end{subfigure}
		\caption{(a) Cumulative distribution of the primary mass of circumbinary planet hosts (blue solid line) compared to a Salpeter initial mass function  \citep{Salpeter:1955fj} with three different $\alpha$ parameters. (b) Two mass-radius relations: Eq.~\ref{eq:mass_radius_relation} from \citet{Eker:2018nw} as a solid black line and an ad hoc creation of a relation for slightly evolved stars as a dashed black line. For comparison the primary and secondary stars in planet-hosting binaries are shown as blue diamonds and red squares, respectively. (c)  Cumulative distribution function of the mutual inclination of the  {\it Kepler} circumbinary systems (black dashed line) compared with a Rayleigh distribution of mutual inclinations with five different $\sigma_{\Delta I}$ parameters.}
\label{fig:results}  
\end{center}  
\end{figure*} 

The binary separations are drawn from a log-normal distribution specified by \citet{Raghavan:2010lr}, where the mean of $\log_{10}T_{\rm bin}$ is 5.03 and the standard deviation of $\log_{10}T_{\rm bin}$ is 2.28, where both values are given in days.

 For the primary masses we draw from a Salpeter initial mass function (IMF) \citep{Salpeter:1955fj} between 0.69 and 1.53$M_{\odot}$. In Fig.~\ref{fig:imf} we show the match between the observed distribution of $M_{\rm A}$ and the Salpeter IMF for three different $\alpha$ parameters: 1.35 and 3.35, with 2.35 being considered standard. Given our uncertainty in $\alpha$, we run simulations for five values within this range.  Secondary masses are calculated based on a uniform distribution of the mass ratio $q$. 

In Fig.~\ref{fig:mass_radius_relation} we show the primary and secondary star masses and radii, and the mass-radius relation in Eq.~\ref{eq:mass_radius_relation} from \citet{Eker:2018nw}. For $M<1M_{\odot}$ this relation works well, for both primary and secondary stars. For more massive stars there is a spread in radius, likely corresponding to some stars evolving within a Hubble time. To test whether having many evolved primary stars would significant affect our results, we create a second, more steep mass-radius relation of $R/R_{\odot}=1.5(M/M_{\odot})^2-0.508$ for $M>1M_{\odot}$. This is chosen in an ad hoc fashion to pass through the largest radius, $R_{\rm A}=1.76R_{\odot}$ for Kepler-38. It must be emphasised that this alternate relation corresponds to an extreme case of many evolved/inflated stars. Its purpose is to test the dependence of the results on the mass-radius relation, and is not to be considered wholly representative of the sample.

The binary eccentricity is taken as circular. Even though this is not realistic for a typical population of binaries, because this paper is concerned with the specific effect of the mass ratio on the observations it is considered reasonable to ignore the binary eccentricity\footnote{A minor exception to this is when calculating the stability limit according to Eq.~\ref{eq:stability_limit}, where we see that the stability limit is a joint function of the binary eccentricity and the mass ratio. However, as shown in Fig.~\ref{fig:stability_limit}, the functional dependence of $a_{\rm crit}$ on $q$ only changes slightly with $e_{\rm bin}$ and typically remains fairly flat.}. The orientation of the binaries is isotropically distributed, meaning a uniform distribution of $\cos I_{\rm bin}$. The initial binary population numbers 20,000,000. Such a large value was necessary given how many detection criteria were applied afterwards.

Circumbinary planets are assigned to each binary with a period drawn from a log-uniform distribution between 0.01 AU and 2 AU. For each planet we calculate the \citet{Holman:1999lr} stability limit (Eq.~\ref{eq:stability_limit}). All circumbinary systems with unstable planets are removed. Roughly 70\% of the original population were removed, largely those with very small semi-major axes. The distribution of the mutual inclination $\Delta I$ is not well-known. In Fig.~\ref{fig:DeltaI} we show the observed distribution compared with Rayleigh distributions with five parameters $\sigma_{\Delta I}$ between $0.5^{\circ}$ and $2.5^{\circ}$. All are consistent with  \citet{Li:2016ng}'s conclusion that $\left<\Delta I\right> \lesssim 3^{\circ}$, but since the transit probability is a sensitive function of $\Delta I$ we run the simulations of all five distributions.

The planet's mass is set to zero since it does not affect its dynamics or detectability. The planet radius is drawn from the \citet{Petigura:2013ur} observed distribution of transiting planet radii from the {\it Kepler} mission. Only planets larger than $3R_{\oplus}$ are considered since this roughly corresponds to the smallest detected circumbinary planet (Kepler-47b), and there is a difficulty in detecting smaller circumbinary planets \citep{Armstrong:2014yq,Martin:2018gr}.

\subsection{Simulating the transiting population}\label{subsec:de-biasing_observed}

 Overall we run 50 simulations, corresponding to five $\alpha$ parameters for the Salpeter IMF for the primary stellar mass, the two mass-radius relations (one standard and one for evolved stars) and five $\sigma_{\Delta I}$ parameters for the Rayleigh distribution of the mutual inclination between the planet and binary orbital planes.

For each simulation, listed here are the steps taken to go from the initial population to the final observed population of circumbinary planets transiting eclipsing binaries.

\begin{enumerate}
\item We create an initial population of 20,000,000 binaries.
\item The binaries are deemed to eclipse or not according to Eq.~\ref{eq:EclipseCriterion}. 
All non-eclipsing binaries are removed from the sample. Binaries with periods longer than two years are cut because any orbiting planets would be very unlikely to transit. Binaries shorter than five days are also removed, because of the observed and theoretically predicted dearth of circumbinary planets orbiting the tightest binaries \citep{Munoz:2015uq,Martin:2015iu,Hamers:2016er,Xu:2016qw,Fleming:2018mf}. Out of the original 20,000,000 binaries, about 60,000 remain after these initial cuts. These binaries are duplicated ten times to allow sufficient statistics after later cuts. This is computationally faster than starting with ten times more binaries originally. 
\item Planets are checked if they have a stable orbit according to the \citet{Holman:1999lr} stability criterion (Eq.~\ref{eq:stability_limit}).
\item Planets are checked if they transit within a {\it Kepler}-like four-year timespan. This comes from calculating the percentage of time spent in transitability according to Eqs.~\ref{eq:Ip_over_time} and ~\ref{eq:Ip_transit_limits}, the nodal precesion period from Eq.~\ref{eq:Tprec2}, and then randomising the start of transitability and seeing if it overlaps with the four-year window. 
\item For all transiting planets the transit depth is calculated using Eq.~\ref{eq:transit_depth_binary}, accounting for dilution from the secondary. A planet is deemed detectable if the transit is deeper than a threshold depth of 0.1\%. This criterion is chosen to corresponds to the smallest circumbinary planet detected to date being $3R_{\rm Earth}$ (Kepler-47b).
\end{enumerate}

\subsection{Results}\label{subsec:de-biasing_results}

\begin{figure*}  
\begin{center}  
	\begin{subfigure}[b]{0.47\textwidth}
		\caption{}
		\vspace{-2ex}
		\includegraphics[width=\textwidth]{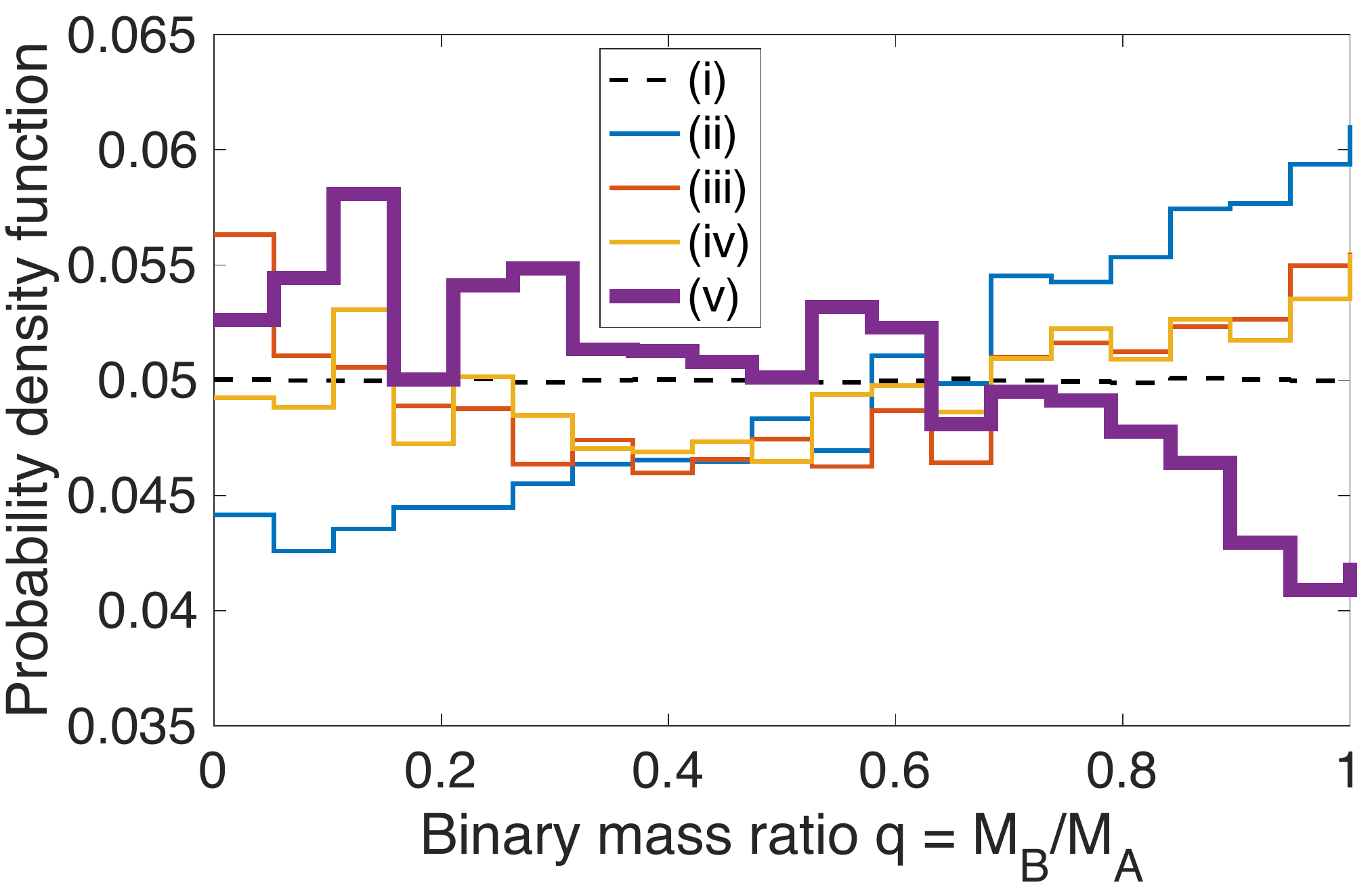}  
		\label{fig:results_histogram}
		\vspace{3ex}
	\end{subfigure}
	\hspace{0.5cm}
	\begin{subfigure}[b]{0.47\textwidth}
		\caption{}
		\vspace{-2ex}
		\includegraphics[width=\textwidth]{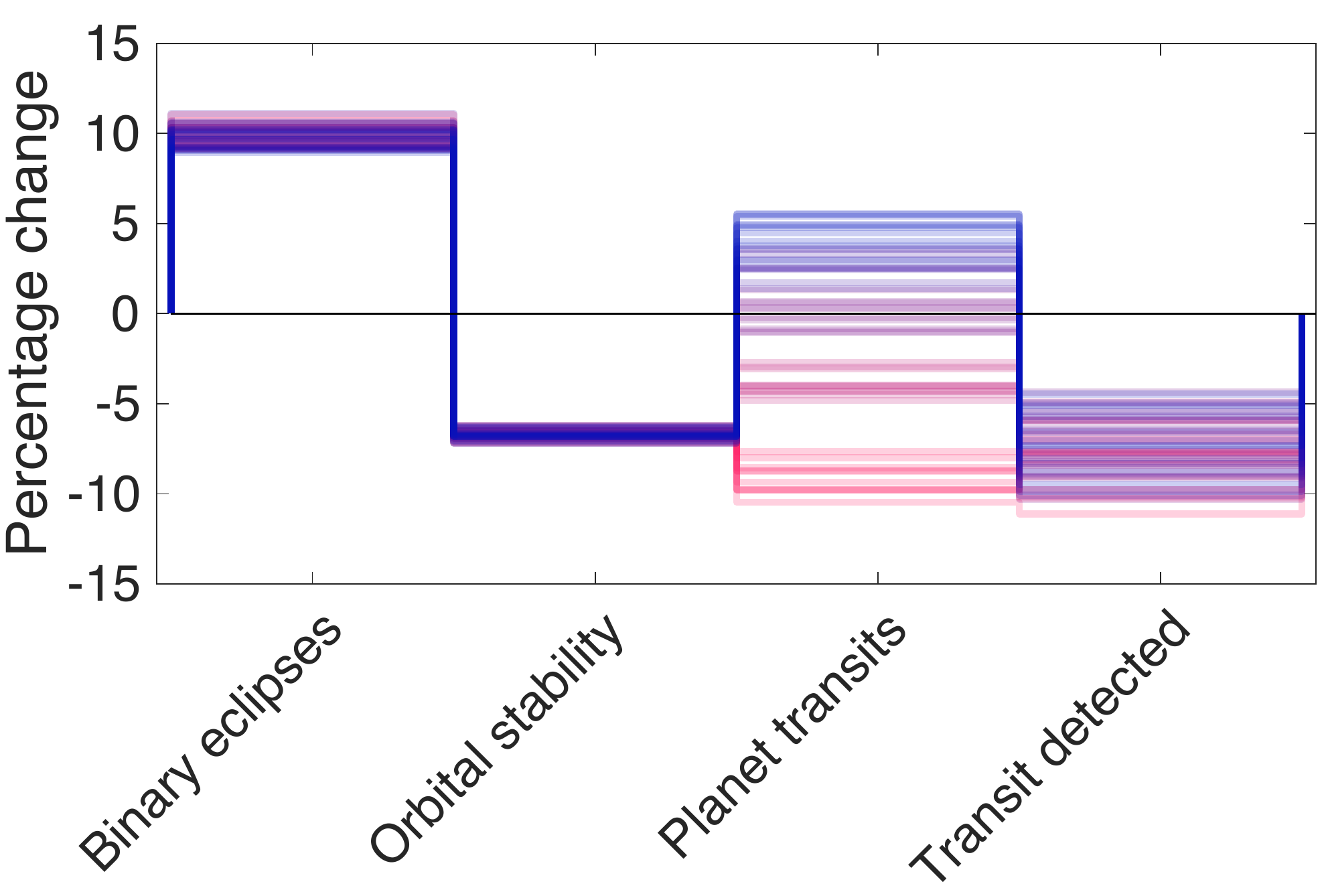}  
		\label{fig:results_changes}
		\vspace{3ex}
	\end{subfigure}
		\caption{ (a) Probability density function of the binary mass ratio distribution for the ``standard'' simulation with Salpeter $\alpha=2.35$, mass-radius relation from Eq.~\ref{eq:mass_radius_relation} and $\sigma_{\Delta I}=1.5^{\circ}$. The original population (i) has a flat distribution (black dashed line). We then apply the four selection criteria from Sect.~\ref{subsec:de-biasing_observed}: (ii) binary eclipses; (iii) orbital stability; (iv) planet transits and (v) transit detected, with the last being the final population (bold purple solid line). This purple curve is our simulated detection bias. (b) Change in the percentage of high mass ratio ($q>0.5$) binaries by each selection effect for all 50 simulations. Results are colour-coded by the simulated $\sigma_{\Delta I}$, from $0.5^{\circ}$ (red) to $2.5^{\circ}$ (blue). Note that this is the step by step percentage change, not the cumulative change. Overall, the results show that the different selection effects largely cancel out, and the transit observations are effectively unbiased as a function of $q$.}
\label{fig:results}  
\end{center}  
\end{figure*} 

In Fig.~\ref{fig:results_histogram} we show for one simulation how the initially flat distribution (i) of $q$ changes after applying each of the four selection effect cuts (ii to v). This may be considered the ``standard'' simulation, with $\alpha=2.35$ for the Salpeter IMF, Eq.~\ref{eq:mass_radius_relation} is used for the mass-radius relation and $\sigma_{\Delta I}$ is used for the mutual inclination distribution, which appears to be the best fit to the observed systems according to Fig.~\ref{fig:DeltaI}. 

Curiously, the final distribution, which is in bold for emphasis, is very similar to the initially flat distribution of $q$. Whilst the different selection effects tend to weight the distribution towards high or low $q$, the net effect is that they largely cancel out. 

Another way of demonstrating this is shown in Fig.~\ref{fig:results_changes}, showing the percentage change of the distribution at each step. We only show the change for high mass ratio $q>0.5$ binaries but the results for $q<0.5$ would be simply mirrored vertically. In this plot results are shown for all 50 simulations. The effect of each individual selection bias can be summarised as:

\begin{itemize}
\item Binary eclipses: high $q$ is always favoured, as expected from the eclipse criterion in Eq.~\ref{eq:EclipseCriterion}. This trend is roughly independent of the simulation parameters.
\item Orbital stability: high $q$ is always disfavoured, as expected from the stability limit in Eq.~\ref{eq:stability_limit}, although this equation is not a monotonic function of $q$. This trend is stringently independent of the simulation parameters.
\item Planet transits: high $q$ is favoured for large values of $\sigma_{\Delta I}$ (blue in Fig.~\ref{fig:results_histogram}) but high $q$ is disfavoured for small $\sigma_{\Delta I}$ (red in Fig.~\ref{fig:results_histogram}). The reasons for this are multi-faceted. For misaligned systems (even just a couple of degrees) the planets are likely to precess in and out of a transit window \citep{Martin:2017qf}. A high $q$ makes the precession faster (Eq.~\ref{eq:Tprec}) and the transit windows longer (Eq.~\ref{eq:Ip_transit_limits}), and hence makes transits more likely. For systems very close to coplanarity, such as with $\sigma_{\Delta I}=0.5^{\circ}$, precession is less important. The bigger factor is that transits are more likely on binaries closer to exactly edge-on ($I_{\rm bin}=90^{\circ}$). Small $q$ eclipsing binaries are more tightly constrained to $90^{\circ}$ due to the small radius of the secondary star, and hence increase the transit probability for near-coplanar planets.
\item Transit detected: high $q$ is always disfavoured, although there is a dependence on the simulation parameters, in particular the mass-radius relation used, because stars with inflated radii significantly reduce the transit depth (Eq.~\ref{eq:transit_depth_binary}).
\end{itemize}

Overall, one selection effect favours high $q$, two favour low $q$ and for one the trend changes as a function of the simulated distribution of $\Delta I$. This means that the individual effects do not add coherently to produce a large skew in the $q$ distribution. Furthermore, all of the effects are only on the order of $\sim10\%$. Therefore, we conclude that transit detections are essentially unbiased with respect to the binary mass ratio.

\section{Comparison with observations of field binaries}\label{sec:observations}

\subsection{Observations of field binaries}\label{subsec:surveys_field_binaries}

\begin{figure}  
\begin{center}  
\includegraphics[width=0.49\textwidth]{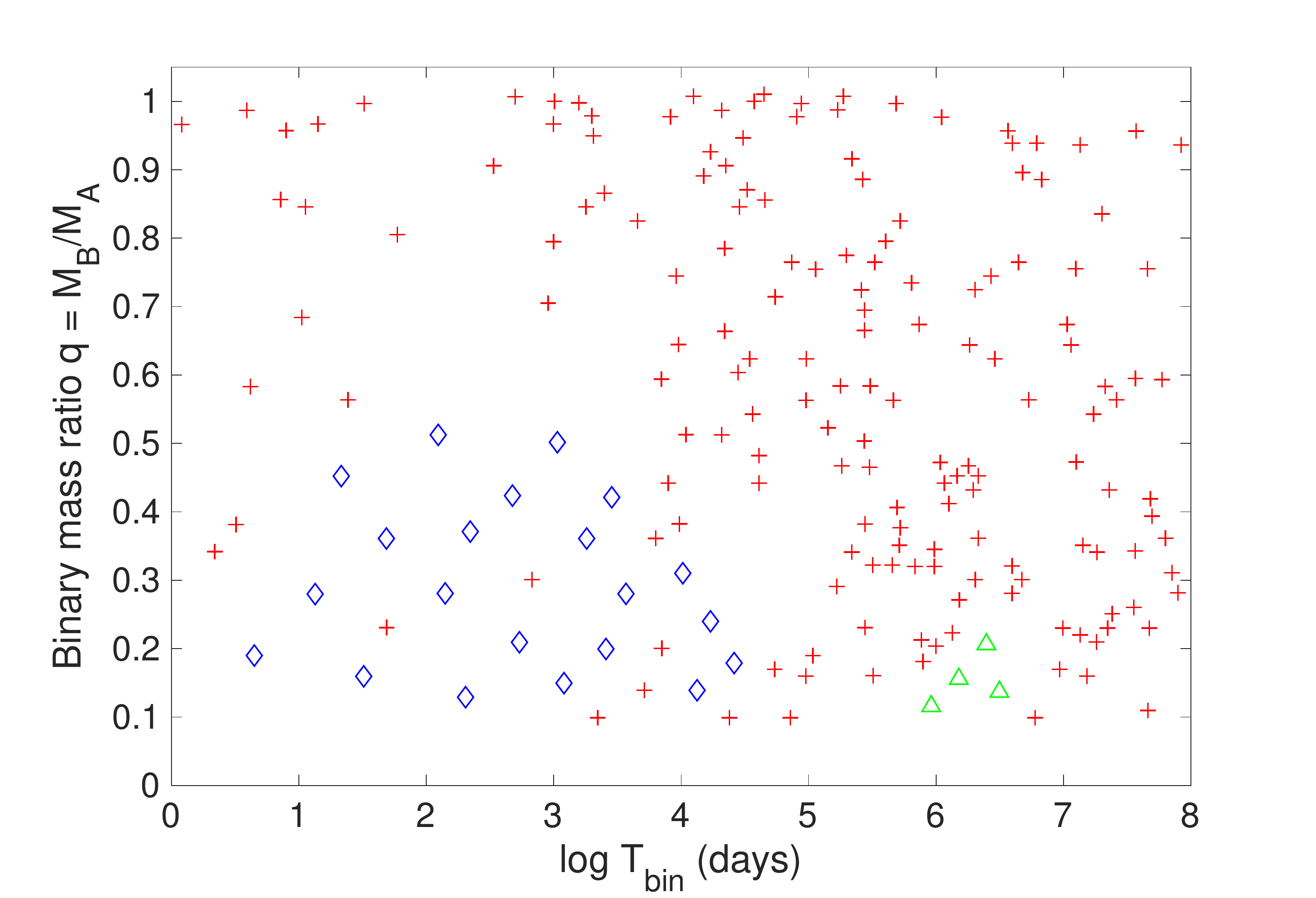}  
\caption{Data of binary mass ratios as a function of the orbital period taken from \citet{Raghavan:2010lr} (red pluses) and completeness corrections made by \citet{Moe:2017nv} (blue diamonds and green triangles). \citet{Moe:2017nv} also made some minor changes to the selection of targets from \citet{Raghavan:2010lr}, particularly with respect to higher-order stellar multiplicity. This figure is a reproduction of Fig. 28 of \citet{Moe:2017nv} with permission from the author.}
\label{fig:moe_maxwell}
\end{center}  
\end{figure} 

The data for the comparison sample comes from the seminal work of \citet{Raghavan:2010lr}, which is a vast collection of over 400 multi-star systems, taken from multiple techniques including radial velocities, imaging and Hipparchos astrometry. By combining different observational methods the paper covers a broad range of  orbits, from tight systems similar to planet hosts to widely separated binaries with periods as long as $10^{10}$ days. These data are shown in Fig.~\ref{fig:moe_maxwell} with a red `+'.

It has been argued that by \citet{Moe:2017nv} that the \citet{Raghavan:2010lr} results suffer from incompleteness issues in two parameter spaces. A small one is for $T_{\rm bin}$ roughly between $10^{5.9}$ and $10^{6.7}$ days and $q\approx 0.1-0.2$. A larger one, which is more significant for our survey, for these relatively close binaries ($T_{\rm bin} \lesssim 10^4$ days). For this period range   \citet{Raghavan:2010lr} uses spectroscopic binaries. However, for small mass ratios the binary is a single-lined spectroscopic binary, and hence the mass ratio cannot be directly calculated, which skews the \citet{Raghavan:2010lr} distribution to high $q$. Within both of these incomplete parameter spaces \citet{Moe:2017nv} fills in the $q$ vs $T_{\rm bin}$ data in \citet{Raghavan:2010lr} by probabilistically deriving synthetic ``observed'' binaries in this parameter space, shown in Fig.~\ref{fig:moe_maxwell} with green triangles and blue diamonds. The main effect of this is to make the distribution of $q$ for short-period binaries more flat, and less skewed towards high $q$.

\begin{figure*}  
\begin{center}  
\includegraphics[width=0.99\textwidth]{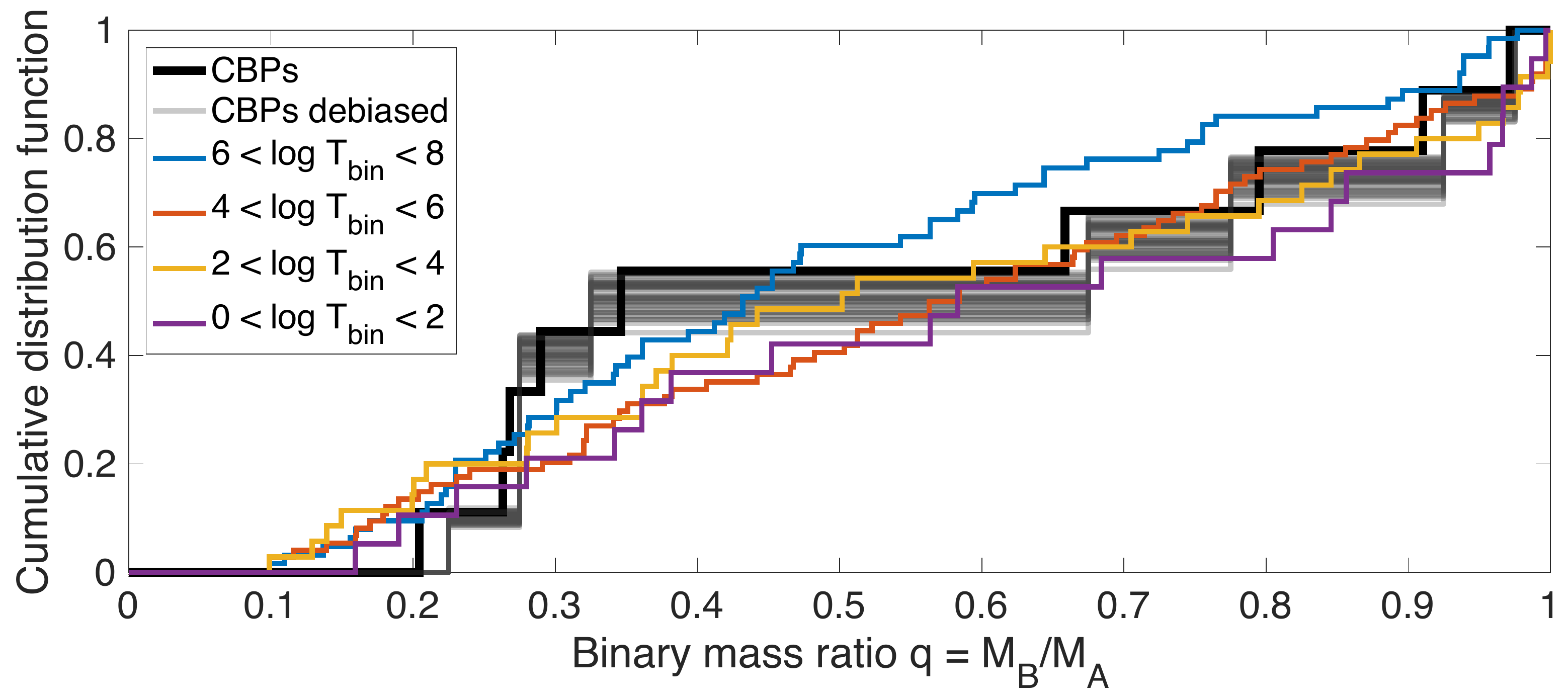}  
\caption{Cumulative distribution functions of the mass ratio of field binaries and circumbinary planet hosts. The planet hosts are plotted as a black solid line for the raw data and grey semi-transparent lines for the de-biased data, for each of the 50 simulations. The data for the field binaries is taken from \citet{Raghavan:2010lr} with application of the completeness corrections from \citet{Moe:2017nv}. The binaries are split into four bins of the binary period, equally sized logarithmically with units of days. }
\label{fig:observations_comparisons}
\end{center}  
\end{figure*} 

In Fig.~\ref{fig:observations_comparisons} we plot the binary data from \citet{Raghavan:2010lr} with the \citet{Moe:2017nv} completeness corrections. The data are separated into four period bins, equally sized logarithmically. We also plot a flat distribution for comparison. The tightest binaries ($T_{\rm bin}=1-100$ days) are  more skewed towards high $q$ than the widest binaries ($T_{\rm bin} = 10^6-10^8$ days), but visually it does not appear to be a significant difference. The tightest binaries have a roughly four times increased abundance of a twins population ($q>0.95$) which is not apparent in wider binaries. \citet{Tokovinin:2014ho} state that the twins excess is even more narrowly confined to $T_{\rm bin}< 20$ days. \citet{Pinsonneault:2006vb} show that in the Small Magelllanic Cloud the percentage of $q>0.95$ systems is as high as 50\% for $T_{\rm bin}<10$ days, but this is for massive stars, not the $\lesssim1.5 M_{\odot}$ stars that host circumbinary planets.

Binaries of all period are shown to be lacking $q<0.1$ companions. This corresponds to the so-called brown dwarf desert \citep{Marcy:2000yf,Sahlmann:2011qy,Kraus:2011ig,Cheetham:2015jr}.

\subsection{Debiased {\it Kepler} circumbinary host sample}

In Fig.~\ref{fig:observations_comparisons} we also plot a cumulative distribution of the mass ratio of circumbinary planet hosts. We plot both the raw data (black solid line) and the de-biased data (black dashed line). The de-biasing is done by multiplying the probability density function of the raw data by the simulated bias curve (i.e. the bold purple curve in Fig.~\ref{fig:results_histogram}). This is done for all 50 simulations, overplotted as semi-transparent grey curves. 

The observed distribution is slightly skewed towards higher $q$ as a result of the de-biasing, but not a substantial amount. There is an overabundance of small mass ratio systems, with roughly 5/9 binaries having $q\in [0.2,0.35]$. On the surface, this appears to be a four times inflated abundance compared to a flat distribution, but such conclusions are likely premature given the poor statistics to date, and if anything the de-biasing process also reduces the strength of this result.

A Kolmogorov-Smirnov test is used to determine if two sets of data are consistent with being drawn from the same distribution. For both the raw and de-biased (from  Fig.~\ref{fig:results_histogram}) $q$ distributions for planet hosts we cannot rule out at $2\sigma$ that they are drawn from the same population as the field binaries, for any binary periods. Interestingly though, at $2\sigma$ a Kolmogorov-Smirnov test similarly cannot rule out that the longest (blue curve) and shortest (purple curve) period binaries are drawn from the population. The poor statistics of the short-period field binary sample are likely the limiting factor, with only 14 discoveries from \citet{Raghavan:2010lr} and an additional 5 synthesised planets from \citet{Moe:2017nv}. Overall, more discoveries of both circumbinary planets and binaries themselves are required to differentiate the populations.

\section{Discussion}\label{sec:discussion}



\subsection{The connected formation and evolution of binaries and planets}\label{subsec:planet_binary_formation}


When binaries form their initial separation is believed to be typically much wider than 1 AU, even as much as hundreds of AU \citep{Bonnell:1994ht,Tohline:2002ln,Kratter:2006qw,Bate:2012rt}. However, the prevalence of much tighter binaries has demanded theories that can shrink an initially wide orbit. Work on this field has existed for decades, however now the discovery of circumbinary planets can shed new light. In brief, there are two main constraints that must be accounted for in any theory of close binary formation:

\begin{enumerate}

\item The tightest binaries, with periods less than $\sim 7$ days, are conspicuously lacking orbiting circumbinary planets \citet{Martin:2014lr,Armstrong:2014yq}. This constraint only applies to planets larger than $\sim3R_{\oplus}$, as observational methods to date have not been sensitive to smaller planets.
\item Slightly longer-period binaries between $\sim 7-41$ days, which we dub ``moderately tight'', host circumbinary gas giants ($R_{\rm P}\gtrsim3R_{\oplus}$) at a rate of $\sim 10\%$ \citep{Martin:2014lr,Armstrong:2014yq,Martin:2019yv}, roughly comparable to planets around single stars. The planets typically orbit almost as close as possible to the binary without being unstable, and on orbits coplanar to within $\sim 4^{\circ}$. Now, based on this paper, we also know that these planet-hosting binaries have mass ratios that, based on current data, are compatible with those of the broader field binary sample.

\end{enumerate}

 For (i), the tightest binaries were already thought to be formed from wider binaries through a process of Kozai-Lidov cycles \citep{Lidov:1961ru,Lidov:1962kx,Kozai:1962qf} under the influence of a misaligned outer third star, followed by tidal friction. There is both theoretical \citep{Harrington:1968oc,Mazeh:1979eu,Kiseleva:1998pe,Eggleton:2006wo,Fabrycky:2007pd,Naoz:2014ti} and observational \citep{Tokovinin:2006la} evidence for this process, although \citet{Moe:2018kl} suggest that there are multiple dynamical pathways to very tight binaries, including Kozai-Lidov cycles both during the main sequence and pre-main sequence phases. This process of Kozai-Lidov cycles was shown to be detrimental to the formation and survival of circumbinary planets \citep{Munoz:2015uq,Martin:2015iu,Hamers:2016er}. A third star may also destabilise planets through evection resonances \citep{Xu:2016qw}.

 For (ii), the formation of slightly wider binaries is yet to be studied in the context of the population of circumbinary planets. Here we encourage and motivate such studies by discussing some of the possible aspects.

There are two leading theories for close binary formation. One is through dynamical interactions in a stellar cluster (e.g. \citealt{Bate:2012rt}), and the other is through accretion-induced migration from a circumbinary disc (e.g. \citealt{Bonnell:1994ht,Kroupa:1995ap,Kroupa:1995bp}). The coplanarity of the circumbinary orbits suggests the accretion disc scenario; circumbinary discs may be expected to be typically coplanar (\citealt{Foucart:2013ys}, although see also \citealt{Martin:2017re,Kennedy:2019qr}) and dynamical interactions with multiple stars could misalign the planet \citep{Munoz:2015uq,Martin:2015iu,Hamers:2016er}. The close-proximity of the circumbinary planets to the stability limit also favours disc accretion, as planets would be unlikely to survive dynamical shrinking of the binary. Problems with the dynamical formation of tight binaries could be overcome if a circumbinary disc were to form after the stellar scattering,  which was seen in some of the simulations of \citet{Bate:2012rt}.  To reproduce the observed planets, the disc would also have to be coplanar with the binary, which may occur on a short re-alignment time-scale \citep{Foucart:2013ys,Pierens:2018fe}. With respect to the binary mass ratios the \citet{Bate:2012rt} simulations predict that the dynamical interactions lead to typically high mass ratio short-period binaries, which is not seen for field binaries or circumbinary planet hosts

 Another argument for the accretion disc theory is the high abundance of circumbinary gas giant planets (similar to that around single stars). This attests to the binary having had a sizable disc, which would have driven both binary migration and planet formation.
 
Studies of accretion discs do have complications though. In addition to shrinking the orbit, the accretion affects the stellar masses and their ratio \citep{Kroupa:1995ap,Kroupa:1995bp}, since the initial proto-binary only has a small fraction of the mass of the parent protostellar cloud \citep{Bonnell:1994ht}. Predictions of how $q$ changes though vary amongst different studies. It may seem intuitive for preferential accretion to the secondary star, since it orbits more closely to the inner edge of what would be a partially truncated circumbinary disc. This would drive $q$ towards unity \citep{Bate:1997df}. However, more recent studies have shown that the evolution of $q$ depends on the temperature of the disc \citep{Young:2015ko,Young:2015gf} or possibly the initial mass ratio \citep{Satsuka:2017mw}. Such disc properties would also effect any embedded planets \citep{pierens:2013kx,Kley:2014rt}. Some studies further complicate matters by predicting that accretion may cause binary orbits  to {\it expand} rather than shrink \citep{Satsuka:2017mw,Munoz:2019qe}.


If a planet has already formed and migrated close to the binary, then any subtle changes to $q$ could have significant effects on the planet. A change in $q$ would change the stability limit, potentially destabilising any planets which migrated perilously close to the binary. The mass ratio also has implications for the stability of planets near mean motion resonances \citep{Henon:1970oh,Dvorak:1989vn,Holman:1999lr,Bromley:2015qy,Quarles:2018ub}.
 
It is also possible that whilst most proto-binaries may form with at wide separations \citep{Machida:2008pg,Bate:2012rt}, some exceptional binaries could form with primordially tight orbits, less than 1 AU. \citet{Machida:2008pg} suggest this could occur as a second phase of collapse from the protostellar cloud. Such binaries may be ideal for planet formation, particularly gas giants, since the truncation of the disc at  $\sim 3a_{\rm bin}$ would be interior to the snow line.

 Finally, one may speculate that the presence of a massive circumbinary planet could somehow alter the evolution of the binary. Given a $\sim 1/1000$ planet-binary mass ratio this may sound outlandish, but there is precedent. \citet{Martin:2015iu} showed that a sub-Saturn-mass planet could actually quench high-eccentricity Kozai-Lidov cycles of a stellar binary, and hence inhibit binary shrinking via tidal friction. However, a comparable mechanism in the paradigm of migration via circumbinary disc accretion is yet to be explored.

Overall, this paper motivates combined theoretical studies of binaries and planets, and suggests that there should not be a strong $q$ dependence on the planet's formation and evolution.

\subsection{Related observational studies}\label{subsec:motivation_observations}

\subsubsection{Mass ratio distribution of the {\it Kepler} eclipsing binary catalog}\label{subsubsec:EB_catalog_mass_ratios}

The mass ratio distribution of the {\it Kepler} eclipsing binary catalog would be the most natural comparison sample for the circumbinary hosts. Unfortunately, this distribution is yet to be derived. The radial velocity survey by \citet{Matson:2017hk} was the first effort, but it only only contains 41 binaries. Furthermore, all of the binaries have $T_{\rm bin} < 6$ days, which is unlikely to be representative of the entire catalog and corresponds to binaries which seemingly do not host circumbinary gas giants. The targets were also selected based on the \citet{Gies:2012ib,Gies:2015qt} search for triple systems using eclipse timing variations,  which may introduce additional biases.

The Villanova eclipsing binary working group\footnote{\url{http://keplerebs.villanova.edu/}} has advertised work done to take radial velocities of the catalog \citep{Kirk:2016sp,Wells:2019rg}. Typically 9-15 measurements will be taken on around 900 of the binaries. However, the radial velocity targets are typically double-lined spectroscopic binaries (Andrei Pr{\v s}a, private comm.). This choice will result in better-characterised binaries, but the mass ratios will be biased to $q\gtrsim 0.5$. So whilst the combined radial velocity and photometric data will yield the the most comprehensive census of short-period binaries to date, the mass ratio distribution will not be entirely comparable with the circumbinary planet hosts.

\subsubsection{Circumbinary discs}\label{subsubsec:ALMA}

Some observations have been made of circumbinary discs, for example L1551 NE \citep{Takakuwa:2012km}, HD 142527 \citep{Boehler:2017ug} and HD \citep{Kennedy:2019qr}. This is a promising new field, for which ALMA will be revolutionary. For now though, the statistics are too sparse to make strong conclusions. Some useful trends to uncover will be the circumbinary disc mass as a function of binary separation and mass ratio, as well as observations of the relative accretion rate onto the primary and secondary stars.

\subsubsection{Radial velocity surveys for circumbinary planets}\label{subsubsec:bebop}


The two largest radial velocity surveys for circumbinary planets are TATOOINE \citep{Konacki:2009lr} and BEBOP \citep{Martin:2019yv}. The two programs have different observing strategies. TATOOINE targets double-lined spectroscopic binaries because the stars can be better characterised and the radial velocity signal by a planet would be measurable in two stars. The binaries are consequently biased to high $q$. BEBOP, on the other hand, targets single-lined binaries, as to avoid the challenging task of deconvolving two overlapping stellar spectra. The BEBOP sample is therefore by construction limited to $q\lesssim 0.4$. Both surveys are yet to yield a planet discovery though.

The analysis of this paper shows that planets exist around binaries of roughly all mass ratios, with a $q$ distribution consistent with being flat. Consequently, even though both TATOOINE and BEBOP target binaries with mass ratios biased in opposite directions, this should not affect the planet-finding ability of either survey.

\subsubsection{The {\it TESS} transit survey}\label{subsubsec:tess}

{\it Kepler's} original mission observed a small patch of the sky continuously for a year, whereas {\it TESS} is observing almost all of the sky but typically in one month blocks  \citep{Ricker:2014qv}. This change in strategy has significant consequences for circumbinary planets. All of the known planets have $\gtrsim 50$ day orbital periods, owing to stability restrictions and an apparent paucity of planets around the tightest binaries. For most {\it TESS} binaries, any surrounding planet would only have a single passing at most.

Kostov et al. (under review) propose a novel technique to find planets which fortuitously transit both stars of an eclipsing binary on a single passing. Combined with primary and secondary eclipses, the four photometric events, with variable transit timing and depths, help classify a circumbinary planet better than a planet transiting a single star once. 

Unlike for {\it Kepler}, which this paper has shown to be effectively unbiased with respect to $q$, this {\it TESS} strategy will be biased towards high $q$ for two reasons. First, the planet is required to have visible transits of both stars, hence demanding a sufficiently massive and bright secondary. Second, a high $q$ binary is also more likely to have observable secondary eclipses, which needed to constrain the binary eccentricity ultimately the planetary orbit.

 The analysis of this paper may be applicable to {\it TESS's} one-year continuous observations at the ecliptic poles if a planet can pass the binary multiple times. It is also likely applicable to ESA's future {\it PLATO} transit survey \citep{Rauer:2014qq}, which will probably have longer baselines than {\it TESS}.

\subsection{Limitations of this work}\label{subsec:limitations}

\subsubsection{Malmquist bias}\label{subsubsec:malmquist_bias}

In a magnitude-limited survey such as {\it Kepler} there is a preferential selection of objects that are intrinsically bright: the Malmquist bias \citep{Malmquist:1922og,Malmquist:1925kb}. A tight and hence unresolved binary will appear brighter than a single star at the same distance, simply owing to the flux contribution of two stars. This becomes more of an effect at higher mass ratios, potentially biasing the eclipsing binary catalog towards such systems. 


\subsubsection{Evolved binaries}\label{subsubsec:evolved_binaries}

Stars more massive than $\sim1M_{\odot}$ may have evolved off the main sequence within a Hubble time, and this would affect our results. This was partially accounted for with a modified mass-radius relation, but we ignore the effects such as mass loss relative to the IMF, changes in luminosity, changes in stellar activity which may make transits harder to detect.

In addition, the evolution of the star into a giant would likely affect any tightly-orbiting circumbinary planets \citep{Kostov:2016hu}. There have been roughly a dozen claims of planets orbiting very tight, post common envelope binaries, such as NN Serpentis \citep{Qian:2009qw,Beuermann:2010fk}, however some questions have been raised over their validity \citep{Zorotovic:2013jd,Hinse:2014rm,Hardy:2015lr,Nasiroglu:2017fg}, with the planets potentially being mischaracterisations of the Applegate mechanism \citep{Applegate:1992ty}.


Finally, \citet{Moe:2017nv} deduced that 30\% of single-lined spectroscopic binaries contain main sequence stars orbited by a white dwarf. This is problematic for the mass ratio distribution deduced from radial velocity surveys of binaries, because $q$ would change over the evolution of one of the stars. For eclipsing binary surveys, the roughly Earth-like radius of white dwarfs would typically  be undetectable and hence this problem would be avoided.

\subsubsection{The by-eye detection of circumbinary planets}\label{subsubsec:by_eye_detections}

 All of the transiting circumbinary planets to date were searched for and discovered by eye. This is because the binary motion and variable planetary orbit  induce transit timing variations that are longer than the transit durations \citep{Agol:2005qy,Armstrong:2013rt},  and hence evade standard detection pipelines. So whilst we can do our best to understand the observational biases of circumbinary planets, it must be conceded that some intangible elements of human detections will remain.


One element which is difficult to quantify is the importance of transits on the secondary star. This is intrinsically connected to the mass ratio. The size of the secondary star affects the transit probability and the luminosity of the secondary star determines whether a transit in front of it will be noticeable in the presence of the much brighter primary. All of the detected planets have transits on the primary star, but only 4/9 have transits on the secondary star. A ``smoking gun'' signature of a circumbinary planet is the large transit timing variations and transit duration variations, both of which can be exposed with a series of primary transits alone. An exception would be the detection of the long-period (1108 days) Kepler-1647b, for which there were only two transits on the primary star and one on the secondary \citep{Kostov:2016yq}. The transit on the secondary star was crucial in constraining the model of the circumbinary system. 


\subsubsection{Small number statistics}\label{subsubsec:statistics}

 We were unable to uncover any statistically significant differences between the distribution of  mass ratio for circumbinary planet hosts and field binaries. One obvious reason for this is small number statistics: 9 circumbinary hosts and 14 field binaries from \citet{Raghavan:2010lr} with $T_{\rm bin}<100$ days. 

However, we recall that studies of the distribution of binary periods for eclipsing binaries with and without transiting planets, the so-called dearth of planets around $T_{\rm bin} \lesssim7$ day binaries, was shown to be different to high statistical significance, yet those claims were made using the same small sample \citep{Martin:2014lr,Armstrong:2014yq}. The difference between this paper and those studies is that they uncovered a {\it strong} observational bias towards finding planets around the tightest eclipsing binaries, and hence their absence was meaningful even for a small sample. In this paper we uncover effectively no bias with respect to $q$. This is not necessarily expected, but the consequence is that the observed spread of $q$ across roughly all possible values is likely indicative of the true distribution.

\section{Conclusion}
\label{sec:conclusion}

 We have investigated the detectability of transiting circumbinary planets as a function of the binary mass ratio $q$. A surprising result is that even though $q$ affects the dynamics and stability of planets, the probability of binary eclipses and planet transits and the dilution of transit depths, these different effects largely cancel out. Overall, transit detections are essentially unbiased with respect to the binary mass ratio.

When applied to {\it Kepler} we show that the distribution of planet-host mass ratios is compatible with the roughly flat distribution of $q$ for field binaries  at any orbital period \citet{Raghavan:2010lr,Moe:2017nv}. What may appear to be a slight over-abundance of planet-hosting binaries with $q\in[0.2,0.35]$ is not statistically significant. 

The mass ratios of close binaries were already believed to be a marker of their formation and evolution, although exactly how is debated. The preliminary result from this paper is that any processes which shape the distribution of $q$ do not drastically effect the formation and evolution of surrounding planets. 

This field of study will be benefitted by new discoveries. Our result that transit detections are unbiased by $q$ works for any transit survey with sufficiently long observing windows such that the planet may pass the binary multiple times. This is applicable to the long-pointing windows of {\it PLATO} and the year-long continuous viewing zones at the ecliptic poles by {\it TESS}.

\section*{Acknowledgements}

I acknowledge funding from the Swiss National Science Foundation. I am very grateful to Dan Fabrycky for thoughtful comments on drafts of this manuscript, and to Moe Maxwell for insight regarding the surveys of binary mass ratios. I also appreciate discussions with Anastasios Fragkos, Georges Meynets, Diego Mu{\~n}oz, Marc Pinsonneault, Lisa Prato, Mads S\o rensen and Amaury Triaud which helped motivate some ideas. Finally, I acknowledge the referee, Hagai Perets, who provided a thorough review, which undoubtedly significantly improved the quality of the paper.

\bibliographystyle{aa}
\bibliography{library_binarymassratios.bib}

\end{document}